\documentclass[11pt]{article}
\usepackage{amsmath}
\usepackage{epsfig}
\usepackage{amssymb}
\usepackage{graphicx}
\usepackage{algorithmic}
\usepackage{subfigure}
\usepackage{url}
\usepackage{wrapfig}
\usepackage[ruled,vlined,linesnumbered]{algorithm2e}
\usepackage{color}
\usepackage{subfigure}
\usepackage[english]{babel}

\makeatletter
\renewcommand{\paragraph}{\@startsection{paragraph}{4}{0ex}%
   {-3.25ex plus -1ex minus -0.2ex}%
   {1.5ex plus 0.2ex}%
   {\normalfont\footnotesize\bfseries}}
\makeatother
\begin{document}

\mathchardef\Gamma="0100
\mathchardef\Delta="0101
\mathchardef\Theta="0102
\mathchardef\Lambda="0103
\mathchardef\Xi="0104
\mathchardef\Pi="0105
\mathchardef\Sigma="0106
\mathchardef\Upsilon="0107
\mathchardef\Phi="0108
\mathchardef\Psi="0109
\mathchardef\Omega="010A

\newcommand{\user}{\mbox{$u$}}
\newcommand{\ulevel}{\mbox{$\bar{u}$}}
\newcommand{\bigsqcap}{\mbox{{\Large $\sqcap$}}}
\newcommand{\dl}{\mbox{$\, [ \hspace*{-1.5pt} [\,$}}
\newcommand{\dr}{\mbox{$\, ] \hspace*{-1.5pt} ]\:$}}
\newcommand{\da}{\mbox{$\, A \hspace*{-6.75pt} A \,$}}
\newcommand{\drightarrow}{\mbox{$\rightarrow \hspace*{-8pt} %%@
\rightarrow$}}

\newcommand{\umodels}{\mbox{$\models_{\ulevel}$}}
\newcommand{\hmodels}{\mbox{$\models_{{\bf H},\bar{u}}$}}
\newcommand{\imodels}{\mbox{$\models_{I,\bar{u}}$}}
\newcommand{\tmodels}{\mbox{\,$\models_{{\bf H}}$\,}}
\newcommand{\db}{\mbox{\,$\langle \Delta, \bar{u}\rangle$\,}}
\newcommand{\tp}{\mbox{\,${\bf T}_{\Delta}^{\bar{u}}$\,}}
\newcommand{\tpi}{\mbox{\,$T_\Delta^{\bar{u}}$\,}}
\newcommand{\md}{\mbox{\,${\bf M}_{\Delta}$\,}}

\newtheorem{defn}{Definition}[section]
\newtheorem{thrm}{Theorem}[section]
\newtheorem{prop}{Proposition}[section]
\newtheorem{lemm}{Lemma}[section]
\newtheorem{obsv}{Observation}[section]
\newtheorem{corr}{Corollary}[section]
\newtheorem{example}{Example}[section]

\newcommand{\vs}{\vspace{1ex}}              % Small vertical space
\newcommand{\vsp}{\vspace{2ex}}             % Vertical space
\newcommand{\vspp}{\vspace{4ex}}            % Large vertical space

\newcommand{\negvs}{\vspace*{-1ex}}         % Small negative vertical space
\newcommand{\negvsp}{\vspace*{-2ex}}        % Negative vertical space
\newcommand{\negvspp}{\vspace*{-4ex}}       % Large negative vertical space

\newcommand{\hs}{\hspace*{1em}}                      % Small horiz. space
\newcommand{\hsp}{\hspace*{2em}}                     % Horiz. space
\newcommand{\hspp}{\hspace*{4em}}                    % Large horiz. space

\newcommand{\ie}{\mbox {\em i.e.}}
\newcommand{\eg}{\mbox {\em e.g.}}
\newcommand{\Eg}{\mbox {\em E.g.}}

\def\blackbox{{\rule{2.2mm}{2.2mm}}}
\def\qed{\hspace*{\fill}\blackbox}

\newcommand{\definitionbox}{\qed}
\newcommand{\examplebox}{\qed}
\newcommand{\lemmabox}{\qed}
\newcommand{\theorembox}{\qed}
\newcommand{\conjecturebox}{\qed}
\newcommand{\algorithmbox}{\qed}
\newcommand{\observationbox}{\qed}
\newcommand{\obsbox}{\qed}
\newcommand{\assumptionbox}{\qed}
\newcommand{\notesbox}{\qed}

\newcommand{\ra}{\mbox{$\rightarrow$}}
\newcommand{\la}{\mbox{$\leftarrow$}}

\newcommand{\fd}{\mbox{$\rightarrow$}}

\newcommand{\munion}{\mbox{$\cup^m$}}
\newcommand{\join}{\mbox{$\bowtie$}}
\newcommand{\distinct}{\mbox{{\tt distinct}}}

\newcommand{\ojoin}{\mbox{${\Join}^o$}}

\newcommand{\xssql}{\mbox{{\bf Xssql}}}

\newcommand{\BA}{\mbox{$\textit{\textbf{a}}$}}
\newcommand{\BB}{\mbox{$\textit{\textbf{b}}$}}
\newcommand{\BC}{\mbox{$\textit{\textbf{c}}$}}
\newcommand{\BD}{\mbox{$\textit{\textbf{d}}$}}
\newcommand{\BE}{\mbox{$\textit{\textbf{e}}$}}
\newcommand{\BF}{\mbox{$\textit{\textbf{f}}$}}
\newcommand{\BO}{\mbox{$\textit{\textbf{o}}$}}
\newcommand{\BP}{\mbox{$\textit{\textbf{p}}$}}
\newcommand{\BQ}{\mbox{$\textit{\textbf{q}}$}}
\newcommand{\BR}{\mbox{$\textit{\textbf{r}}$}}
\newcommand{\BV}{\mbox{$\textit{\textbf{v}}$}}
\newcommand{\BL}{\mbox{$\textit{\textbf{l}}$}}
\newcommand{\BI}{\mbox{$\textit{\textbf{i}}$}}
\newcommand{\BH}{\mbox{$\textit{\textbf{h}}$}}
\newcommand{\BS}{\mbox{$\textit{\textbf{s}}$}}
\newcommand{\BK}{\mbox{$\textit{\textbf{k}}$}}
\newcommand{\BT}{\mbox{$\textit{\textbf{t}}$}}
\newcommand{\BX}{\mbox{$\textit{\textbf{x}}$}}

\begin{center}
{\Large {\bf A Novel Model for Distributed Big Data Service Composition using Stratified Functional Graph Matching}}\\
{\bf Carlos R. Rivero}$^\natural$ and {\bf Hasan M. Jamil}$^\flat$\\
$^\natural$Department of Computer Science, Rochester Institute of Technology, USA\\
{\tt crr@cs.rit.edu}\\
$^\flat$Department of Computer Science, University of Idaho, USA\\
{\tt jamil@uidaho.edu}
\end{center}

\begin{abstract}
A significant number of current industrial applications rely on web services. A cornerstone task in these applications is discovering a suitable service that meets the threshold of some user needs. Then, those services can be composed to perform specific functionalities. We argue that the prevailing approach to compose services based on the ``all or nothing'' paradigm is limiting and leads to exceedingly high rejection of potentially suitable services. Furthermore, contemporary models do not allow ``mix and match'' composition from atomic services of different composite services when binary matching is not possible or desired. In this paper, we propose a new model for service composition based on ``stratified graph summarization'' and ``service stitching''. We discuss the limitations of existing approaches with a motivating example, present our approach to overcome these limitations, and outline a possible architecture for service composition from atomic services. Our thesis is that, with the advent of Big Data, our approach will reduce latency in service discovery, and will improve efficiency and accuracy of matchmaking and composition of services.
\end{abstract}


% We no longer use \terms command

\section{Introduction}

Since the advent of the World Wide Web, a significant number of companies have relied on web services to expose and share their products and services, and manage their activities. One of the most important tasks in current industrial applications that use web services consists of discovering suitable services that meet some threshold of the user needs, which is usually referred to as service matchmaking~\cite{journals/ws/SycaraPAS03}. The expected output of this task is a ranking of available services that meet the matching threshold of the user needs sorted by relevance. In the event a single service is not available that atomically satisfies the user needs, multiple atomic services are usually strung together to design a composite service that amounts to a process known as service composition~\cite{conf/smc/CuzzocreaF11}.

In the context of Big Data, it can be argued that in addition to the size of the data, the variability and the number of candidate repositories or services pose a significant challenge for users toward efficiently finding and using them. We posit that the sheer number of candidate services make it necessary to automate the process of service matchmaking and composition since a large number of services may potentially satisfy user needs~\cite{conf/semweb/RaoS04}, often partially.

While atomic service matching or composing atomic services  based on the ``all or nothing'' model~\cite{conf/smc/CuzzocreaF11} covers many applications, services requiring components of two or more web services cannot be composed currently due to technical hurdles. We believe disallowing partial service matching, and service composition using such partial matchings limit the applicability of service composition in real-world scenarios, since they lead to exceedingly high rejection of potentially suitable services. Allowing ``mix and match'' of parts of services will enlarge the class of applications that can be modeled or the services that can be supported, i.e., instead of matching a single service, it is possible that the best solution to some user needs may entail the composition of parts of several services in a meaningful way.

In this paper, we present a new approach to service matchmaking and composition that, to the best of our knowledge, is the first that addresses distributed, fine grain and approximate service discovery and composition in the context of Big Data. In our approach, we represent services as graphs, and use ``graph summarization" to describe them at multiple layers. We then use the resulting stratified graphs to match services at different description granularity to improve precision and recall. While we view the graph matching technique based on stratified graph summaries and the issue of graph summarization as an orthogonal research and leave them outside the scope of this paper, we believe that developing such algorithms are possible along the lines of~\cite{ShoaranTW13,DoanAL12,LiuYC12,ZhangTP10,LeFevreT10}, though not trivial. Furthermore, we also propose to allow composing services from partial matches of services and stitch them together to architect a new service. Such an approach will allow composing parts of atomic services not possible in contemporary models.

As a motivation for our approach, we use an example from the benchmarking domain to illustrate the salient features of our service matching model. We also use this example to discuss the limitations of the existing models and show how we can be overcome them. We also briefly present a possible representation model for service description that will allow partial matching and our proposed ``service stitching" as we envision it. While we are agnostic about the standards to choose, we use features borrowed from BPMN and OWL-S to represent service description workflow graphs for the time being. Finally, our thesis is that, our approach will effectively reduce latency in service discovery, improve efficiency and accuracy of service matchmaking and composition by taking different layers of abstraction into account, and by composing best-fitting services from disparate services.

The remainder of the paper is organized as follows: related research is presented in Section~\ref{sec:related-research} followed by a motivating example in Section~\ref{sec:motivating-example} where we discuss the merits of our approach in relation to the limitations of current approaches. In Sections~\ref{sec:service-matchmaking} and~\ref{sec:service-composition} respectively we describe our idea of service matchmaking and composition. We wrap up our presentation by presenting our overall conceptual model in Section~\ref{sec:model} before concluding in Section~\ref{sec:conclusions}.

\section{Related Research}
\label{sec:related-research}

Sycara et al.~\cite{journals/ws/SycaraPAS03} successfully demonstrated that including semantic annotations over services improve service matchmaking and composition. In semantic web, several models such as OWL-S~\cite{conf/www/MartinBMMPSMSS07} and WSMO~\cite{conf/esws/VitvarKVF08} support such annotations. Interestingly, many service matchmaking and composition approaches that use semantic annotations are based on graph matching~\cite{conf/www/KusterKSK07,journals/ws/KluschFS09}, and an upward trend in leveraging graph matching is clearly visible (e.g., \cite{MaL12,conf/smc/CuzzocreaF11}).

Graph matching currently is a very active area of research which can broadly be classified into two groups: exact and subgraph matching, and approximate matching. Exact matchers aim to find those structurally isomorphic subgraphs that exactly match a specific query. Some of the prominent systems include GraphQL~\cite{conf/sigmod/HeS08}, Turbo$_{\mbox{iso}}$~\cite{conf/sigmod/HanLL13}, and NetQL~\cite{conf/icde/RiveroJ14}. Approximate graph matching, on the other hand, focuses on finding those subgraphs that approximately match a specific query within a degree of deviation. Such matching are useful when the query graph is tentative, or data sets are noisy or encomplete. Some of the well regarded systems in approximate matching include TALE~\cite{conf/icde/TianP08}, SAPPER~\cite{journals/pvldb/ZhangYJ10} and TraM~\cite{journals/tcbb/AminFJ12}.

Currently, a new direction of research in graph summarization is also taking root \cite{KoutraKVF14,ShoaranTW13,DoanAL12,LiuYC12,ZhangTP10,LeFevreT10} that has significant potential in service matchmaking as we advocate in this paper. While the main focus remains on summarization of graphs based on types (node or edge type), efforts are also underway to summarize graphs to discover functional similarity or equivalence (e.g., \cite{MaR07}). Such functional summarization of graphs have been instrumental in designing equivalent workflow design policies for commercial applications \cite{LiangL07}, and can be a powerful tool for service matching based on functional equivalence \cite{PaliwalSVXA12,HumaGEJ12}.

While the logic based service description representation (e.g., \cite{journals/ws/KluschFS09,journals/ws/KluschFS09}) are different than the graph based representations in OWL-S (e.g., \cite{conf/smc/CuzzocreaF11}), neither are flexible enough to accommodate errors or closeness in form of approximate matching. In real life applications, where large number of diverse services are possible, it is difficult to imagine user queries to deterministic in terms of its needs with respect to available descriptions of services. Most likely scenario is that most services will fail to match when an expectation (a query) is submitted and several components of multiple services when strung together may satisfy that need. In this paper, we take the position that an approximate graph matching over stratified summarized graphs of web service descriptions will yield a more realistic and effective service discovery that will allow service stitching to orchestrate a target service from multiple partially matched services. In the remainder of the paper we introduce such a model.

\section{A Motivating Example}
\label{sec:motivating-example}

Let us consider an application where a user wishes to study the performance of a number of query engines for benchmarking, each of which has an entry point by means of which queries can be issued. In addition to the entry points, it is necessary to use a set of queries to test the engines. These queries require data that can be given as input or can be randomly generated according to a given schema. During the benchmarking task, the performance of each query engine is measured based on a given performance variable, such as the time used or memory consumed by the engine to execute the queries over the data. Finally, the expected output is a ranking in which the engines are sorted according to their performance. In our recent research, we have actually successfully applied such an approach to benchmark SPARQL engines over RDF data in~\cite{journals/tkde/RiveroHRC13}, where we devised a statistical evaluation methodology to compare those engines side by side. In this paper, we leverage an abstraction of this approach to focus on the benchmarking of relational engines by means of SQL queries as follows.

\begin{figure}[ht]
\centering
\includegraphics[scale=0.85]{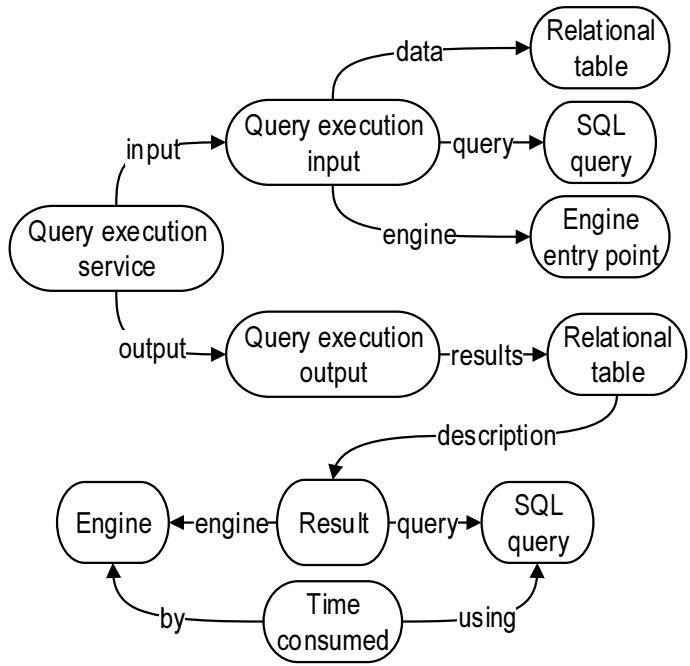}
\caption{Sample model of the Query execution service.}
\label{fig:query-exec-model}
\end{figure}

\begin{figure}[ht]
\centering
\includegraphics[scale=0.85]{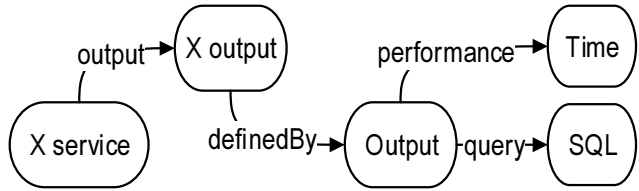}
\caption{Sample model of the user needs.}
\label{fig:query-exec-output-user-needs}
\end{figure}

In this example, assume that there exists a service that takes a relational table, a set of SQL queries and a set of engine entry points as input, and it outputs a relational table that contains the time consumed by each query engine to execute the queries over the input data. Figure~\ref{fig:query-exec-model} shows an example of a model to describe the inputs and outputs of this service called the {\em query execution service}. If a user wishes to use such a service, Figure~\ref{fig:query-exec-output-user-needs} shows an example of a model that describes the user needs. Note that the query topology is different from the description topology, and the name of the nodes and edges are not the same. Due to these dissimilarities, current service matchmakers will reject the {\em query execution service} as a candidate match. The main reasons being a mismatch in topology and semantics, which is extremely difficult to resolve using existing technologies including approximate graph matching \cite{conf/smc/CuzzocreaF11}.

To overcome these limitations, it is necessary to take into account the semantic and topological disparities for a successful matchmaking. For instance, if it is possible to partially match the semantics of a relational table, the system should output such a service as a potential candidate with a estimate of semantic closeness. Although the matching process will be involved and significantly complicated, but doing so will improve the matching quality, precision and recall.

Once we allow flexible matching, we can consider allowing heterogeneity, and a ``service stitching" paradigm in which we allow services to be composed by accepting partial services from different sources. For example, let us assume that the input data format user wants to use for the {\em query execution service} is in text format while the service itself requires it in tabular format. This mismatch could be easily resolved by using an auxiliary service for format conversion without rejecting the match. This can be made possible if the semantic descriptions for the services (input output behavior of each of the nodes in the service description graphs) are available, and if we allow plugging in external components into a service.

Finally, once we allow external plug-ins, we should be able to actually generalize this concept to allow extracting partial services from a description, and stitch several such partial descriptions to compose an entirely new service to meet a user need. Though challenging, such an approach will allow orchestrating services that does not exist even though the components needed do. The idea is detailed in the remainder of the paper in detail.

\section{Functional Graph Summarization for Service Matching}
\label{sec:service-matchmaking}

\begin{figure}
\centering
\subfigure[Visualizing the performance of query engines using random data.] {
    \label{fig:random-data-wf}
    \includegraphics[scale=0.65]{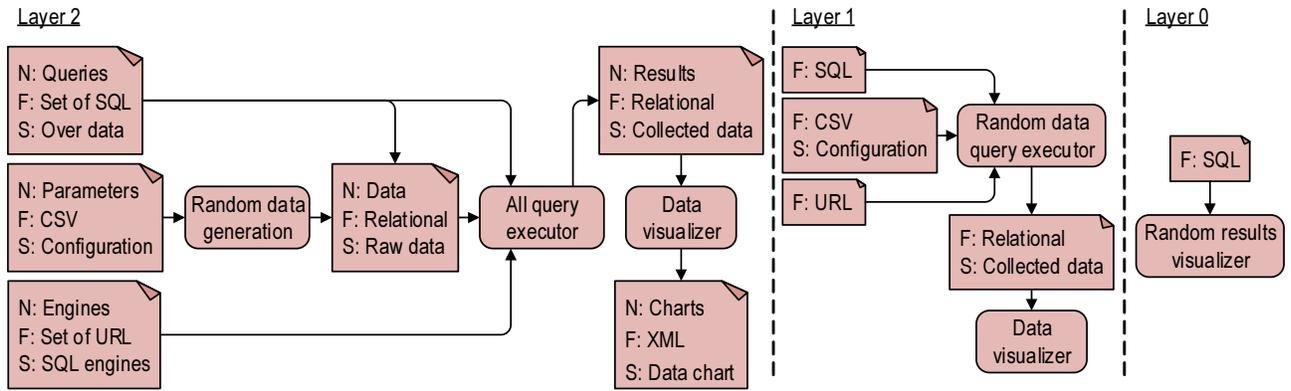}
}
\\
\subfigure[Ranking of query engines using Wilcoxon's signed-ranked test.] {
    \includegraphics[scale=0.65]{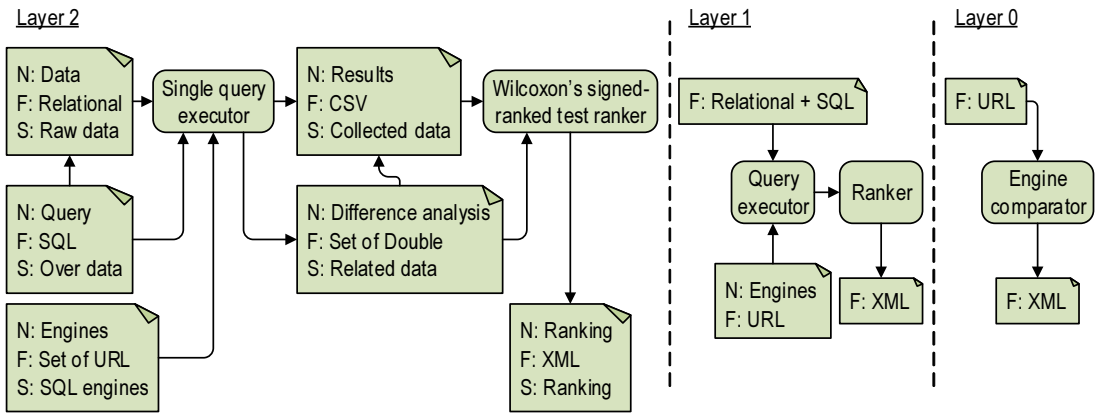}
    \label{fig:wilcoxon-wf}
}
\\
\subfigure[Comparison of samples using Student's t-test.] {
    \includegraphics[scale=0.65]{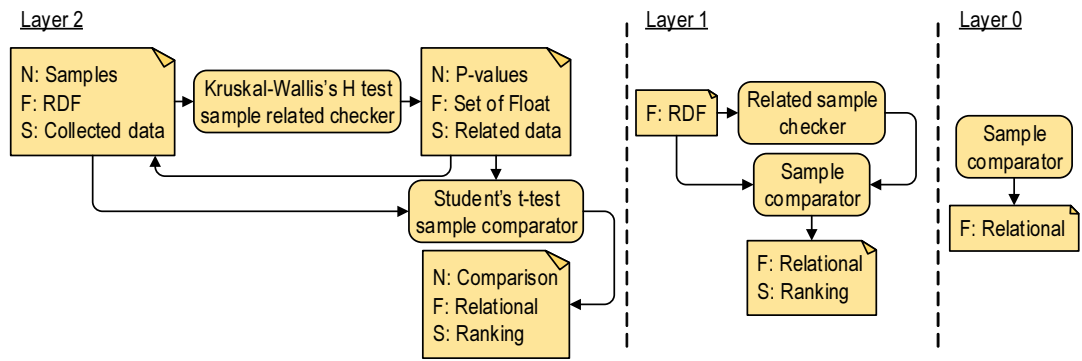}
    \label{fig:student-wf}
}
\caption{Examples of composite services and their summarizations.}
\label{fig:wf-examples}
\end{figure}

We illustrate our approach to flexible service matching using a high level abstraction of the example of relational query engine benchmarking in \cite{journals/tkde/RiveroHRC13}. In which, the process of benchmarking has two critical steps: The first step consists of executing a set of queries over a supplied data set  using the engines being tested. In the second step, we analyze the performance of the engines, so we need a method to compare the values of the performance variable that we obtained in the previous step. We consider these values as samples of an unknown random variable, so we use Kruskal-Wallis's {\em H} test~\cite{HollanderW99} to determine if there are statistically significant differences among the performance variable for the engines under test. If there is no performance difference among the engines, it means that they all behave statistically identically with respect to the performance variable. Otherwise, we use Wilcoxon's signed rank test~\cite{HollanderW99} to rank the engines according to their performance. Taking inspiration from this methodology, we devised the three composite service descriptions represented in the form of graphs in Figure~\ref{fig:wf-examples}.

Figure~\ref{fig:random-data-wf} presents a composite service whose goal is to visualize the performance of a set of query engines based on some randomly generated data. In this figure, atomic services are represented as oval rectangles, and input and output data are represented as rectangles with folding corners. In this representation, data nodes are shown to have three fields: 1) the name (N) of the input/output data; 2) the format (F) of the data, which can be CSV, relational, XML, or RDF, to mention a few; and 3) the semantics (S) of the data, which describes the contents of the data node. We then use a set of pre-defined labels to describe the functionalities of these graphs and their components as follows.

Let us assume that we have a distinguished set of labels $\cal L$. Each label $l\in \cal L$ represents a well understood concept in real life, such as ``Student's t-test", or ``Ranking". To be able to associate a concept with a graph that captures its spirit, we use the function $\Gamma:{\cal L} \mapsto 2^\mathbf{G}$, where $\mathbf{G}$ is a set of graph descriptions of the form $\langle G, D\rangle$ such that $G$ is a graph and $D$ is its input output behavior described in terms of data, format, analytical tools and other needs for the graph to function according to its specification (as shown in Figure~\ref{fig:wf-examples}). Note that a concept may map to a set of such graphs that are likely implementations of that concept. The selection of anyone of these graphs largely depends on how well the input output behavior of the graph matches with the expected behavior. The inverse function $\Sigma:\mathbf{G} \mapsto {\cal L}$, on the other hand, associates with each graph description a conceptual label that describes it uniquely.

We say, a graph description corresponding to a service is {\em atomic} when no sub-graph of it are mapped to a conceptual label in $\cal L$, i.e., a description is self-contained and indivisible. Otherwise, it is called {\em composite} -- composed of atomic and other composite descriptions. For example, the (composite) service in Figure \ref{fig:random-data-wf} comprises of three atomic services: 1) {\em Random data generation} that automatically generates a set of random relational data based on a set of input parameters. 2) {\em All query executor} which takes a number of SQL queries, a set of relational raw data and a set of query engine endpoints as input, and outputs the performance of the engines\footnote{Note that the queries and the data must be related, which is represented as an edge connecting both data nodes. Also note that the engine endpoints are modeled as URLs.}. 3) The {\em Data visualizer} is responsible for plotting the performance results in a set of charts. These atomic services are at the lowest level of representation of the composite service in Layer 2 in Figure~\ref{fig:random-data-wf}. Additionally, this service is summarized into two different layers of abstraction. In Layer 1, {\em Random data generation} and {\em All query executor} have been grouped into a single composite service called {\em Random data query executor}. In other words, the data nodes have been summarized containing their formats and some of their semantics. In Layer 0, the whole service has been summarized into two nodes: a data node that represents some input data in SQL format, and another composite service called {\em Random results visualizer}.

\begin{figure}
\centering
\subfigure[Query executor and ranker.] {
    \includegraphics[scale=0.90]{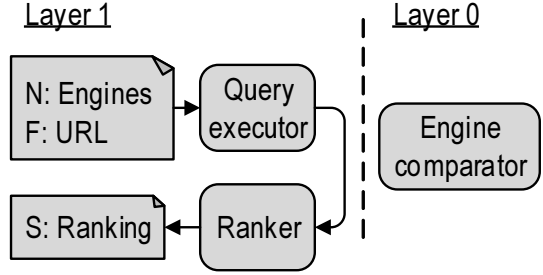}
    \label{fig:eng-cmp-high}
}
\\
\subfigure[Execute queries, analyze the differences and rank them.] {
    \qquad
    \includegraphics[scale=0.90]{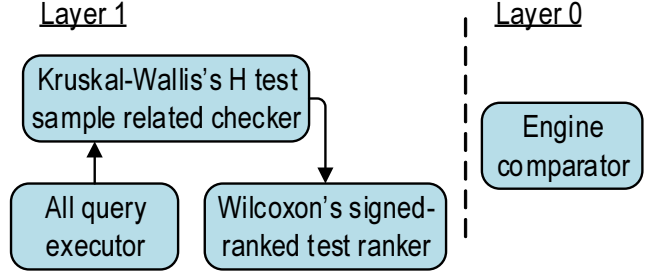}
    \qquad
    \label{fig:eng-cmp-low}
}
\caption{Sample queries of user needs.}
\label{fig:user-needs}
\end{figure}

Figures~\ref{fig:wilcoxon-wf} and \ref{fig:student-wf} show two similar services at three different levels of summarization. The first of which implements the service in Figure \ref{fig:random-data-wf} using Wilcoxon's signed-ranked test, while the second implements it using Kruskal-Wallis's H test and Student's t-test. Now assume that a non-expert user expresses her need to benchmark a set of query engines and rank them according to their performance as the query graph shown in Figure~\ref{fig:eng-cmp-high}, expressed equivalently at two different layers of detail or summarization. To find a matching service, we can contemplate using an approximate graph matcher to identify the candidate services that match with this query graph and the degree to which they match. Regardless of the query at Layer 1 or Layer 0, we can expect a matcher to fail or return a poor match with any of the three services in Figure \ref{fig:wf-examples} due to vastly different topology or labels. But a more smarter matcher that takes into account the semantic meaning of the labels, and is able to consider summarized concepts will be able to recognize a possible match. For instance, this query will not match well with the service in Figure~\ref{fig:random-data-wf} at Layer 0, but it will partially match if we compare Layer 1 of the query with Layer 2 of the service, i.e., both contains a {\em Query execution service}. With respect to the service in Figure~\ref{fig:wilcoxon-wf}, the query will partially match at Layers 0 and 1, i.e., {\em Engine comparator} node, and {\em Query executor} and {\em Ranker} nodes, respectively. Regarding the service in Figure~\ref{fig:student-wf}, it is possible to match the single node at Layer 0 ({\em Sample comparator} and {\em Engine comparator}). At Layer 1, it will partially match with the output nodes since both are rankers. The main observation here is that the query should not fail, and we should be able to find a match despite the disparity in the details of the service description, or the query posed.

If we now consider the queries in Figure \ref{fig:eng-cmp-low}, in which the user specifically wants to use Kruskal-Wallis's H test to analyze the differences between the engines, and use Wilcoxon's signed-ranked test to rank the engines according to their performance, we quickly discover that regardless of version of the query in Layer 1 or Layer 0 is submitted, it will fail. In this case, the query in Layer 1 will partially match with the service in Figure~\ref{fig:random-data-wf} at Layer 2, and with the service in Figure~\ref{fig:wilcoxon-wf} at Layers 0 and 1. Furthermore, while the service in Figure~\ref{fig:student-wf} will not match at Layer 0 with the query, but the query at Layer 1 will match with the Layer 2 description of the service. Thus, it is important to note that while no service matches the query enough in Figure~\ref{fig:eng-cmp-low}, it is possible to orchestrate a service by combining components of different services in a meaningful way, which we advocate in this paper.

\begin{figure*}
\centering
\includegraphics[scale=0.75]{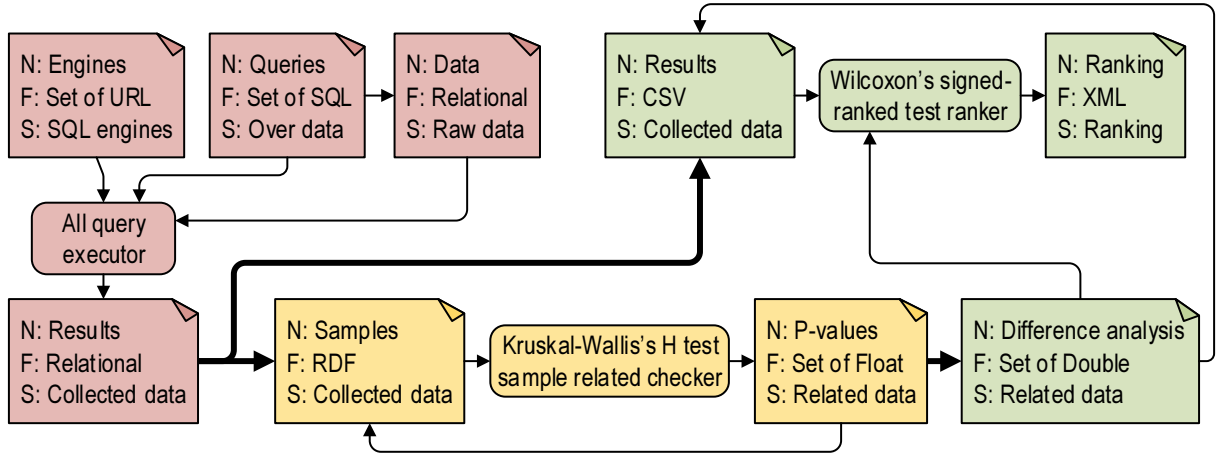}
\caption{Example of a best-fitting composite service (thick arrows represent format conversions among the data).}
\label{fig:final-composite-service}
\end{figure*}

\section{Service Stitching}
\label{sec:service-composition}

Our main goal is to gain the ability to match or orchestrate services with respect to a query described with well understood concepts in the form of a specialized graph from a large set of service descriptions, also represented using graphs. In contrast to most existing approaches, we allow service descriptions to be as detail as possible and at the discretion of the designer and service provider but aim to compensate for any omission in detail to return a successful match. This approach removes the limitations imposed by current systems in two ways.

First, the topology of the service description graphs need not match the query graph structurally so long we are able to match them conceptually at different summarization levels using the functions $\Gamma$ and $\Sigma$. For this to be successful, we primarily insist upon a well understood vocabulary of concepts, and a mechanism to roll-up and drill-down the summarization hierarchy of the description and query graphs to find a match, using a process we call {\em stratified graph matching}. As we have mentioned at the outset, we consider research in stratified graph matching to be orthogonal to our presentation here and needs further investigation. The advantage afforded by this approach is particularly liberating for the service providers. For instance, the disigner of the service in Figure~\ref{fig:student-wf} can now choose to expose it at any level and not worry about its possible match with a query. Instead, she is now free to focus on the details on how best she can describe the service as one single description.

Second, service discovery based on stratified graph matching also inspired a novel approach to service orchestration at the finest granularity, we call {\em service stitching}. Stacking up multiple services in a pipeline to compose a complex service manually is not new. What is new in service stitching is that a query can now be directly mapped to multiple services using the same principle of stratified graph matching because we are allowed to look for services elsewhere in a piecewise fashion, in this instance laterally. With a slight engineering, we can also allow partial services in a description to be invoked.

For example, consider again the query shown in Figure~\ref{fig:eng-cmp-low} consisting three concepts that match with the atomic services in Figures~\ref{fig:random-data-wf},~\ref{fig:wilcoxon-wf}, and~\ref{fig:student-wf}, respectively, but not as single service. We can visualize connecting the output of the Kruskal-Wallis's H test sample checker service in Figure~\ref{fig:student-wf} with the input of the Wilcoxon's signed-ranked test ranker service in Figure~\ref{fig:wilcoxon-wf}, since the former outputs the difference analysis and the latter takes this analysis as input, i.e., the output/input data share the same semantics. The ability to stitch partial services to construct a target service will hinge upon our ability to enter and exit a desired points in a service description graph, a practice not allowed in current service implementations. However, in our approach, all graph descriptions include input out behaviors and thus includes the building blocks needed to support service stitching as detailed in the following section.

\begin{figure}
\centering
\includegraphics[scale=0.85]{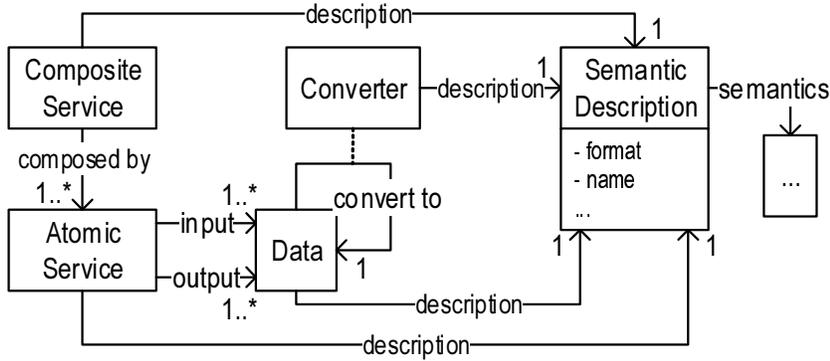}
\caption{Conceptual model.}
\label{fig:conceptual-model}
\end{figure}

\begin{figure*}
\centering
\includegraphics[scale=0.75]{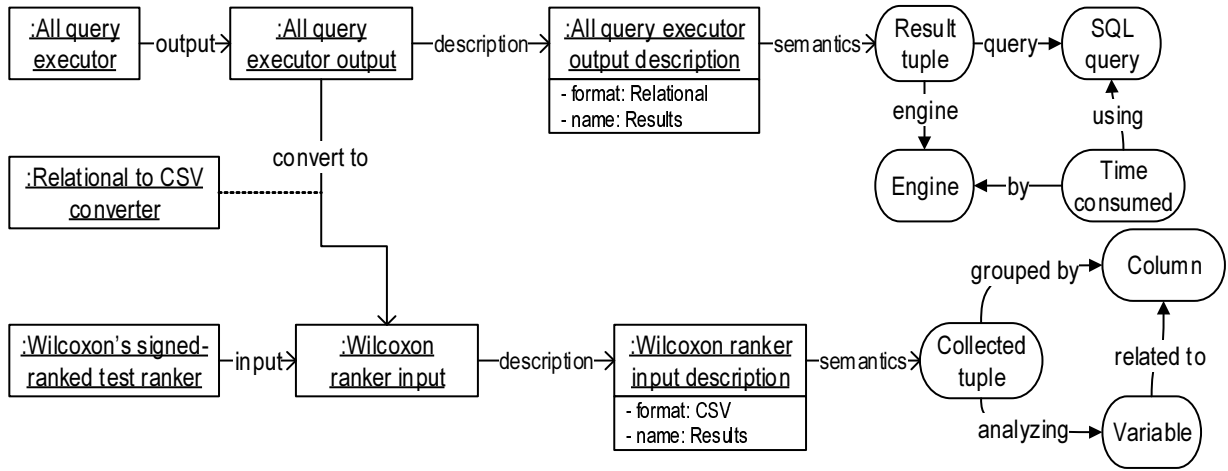}
\caption{Sample instantiation of our conceptual model.}
\label{fig:sample-instantiation}
\end{figure*}

\section{Conceptual Model}
\label{sec:model}

We hinted on stitching Kruskal-Wallis's H test with the Wilcoxon's signed-ranked test ranker services in the previous section. Since the Kruskal-Wallis's H test sample checker in Figure~\ref{fig:student-wf} expects the input samples in RDF format, and the All query executor service in Figure~\ref{fig:random-data-wf} outputs the data in relational format, a suitable conversion service is required to make them compatible. Once we are able to stick in such conversions, we are able to orchestrate the service in Figure~\ref{fig:final-composite-service} (format conversions shown as thick arrows). Therefore, we assume that we have a database of converter services is available to facilitate simple heterogeneity resolution, e.g., from relational data to RDF~\cite{conf/www/AuerDLHA09}, or from XML files to relational data~\cite{journals/is/AtayCLLF07}, to mention just a few.

To support the system we envisioned in the previous sections, a new conceptual model for service description and matching is required. Figure~\ref{fig:conceptual-model} presents an UML-like conceptual model that can be used in this context, which allows to represent that a composite service is composed of a number of atomic services. Each atomic service has a number of inputs and outputs, which can be automatically transformed by means of converters. Finally, each composite service, atomic service, input/output data, and converter is related to a semantic description that allows us to automatically perform the matchmaking and composition of atomic services.

It is important to note that our model is agnostic to the implementation of the system, i.e., it can be used in conjunction with BPMN or OWL-S processes to describe workflows. In this context, our approach has similarities with the description of processes in OWL-S, for instance, atomic, simple and composite processes in OWL-S can be seen as our service descriptions at Layer 2, the summarization of the services in the different abstraction layers, and our composite services, respectively. Furthermore, OWL-S relies on bindings to specify how the atomic processes are connected, which is also similar to our specification on how to connect two or more atomic services.

For the semantic descriptions of the atomic services in our conceptual model, it is possible to use any of the existing approaches in the literature, such as OWL-S or WSMO. We have included in our conceptual model the Name, Format and Semantics fields that we have used in previous sections to motivate our approach; however, this is a clear point of variability of our conceptual model since, depending on the scenario, it is necessary to use different semantics, and it could be suitable to use one specific approach to add semantic annotations to the services.

To illustrate the use of our conceptual model, we present a sample instantiation in Figure~\ref{fig:sample-instantiation}. In this instantiation, we have used OWL-S to model the semantics of the input/output data. It corresponds to a part of the composition of atomic services in Figure~\ref{fig:final-composite-service}, specifically, how we transform from the output data of the All query executor service to the input data of the Wilcoxon's signed-ranked test ranker service, which are represented as a relational table and a CSV file, respectively. Note that the Relational to CSV converter has access to the semantic descriptions of both the relational table and the CSV file, which are used to perform the conversion between them.

\section{Conclusions}
\label{sec:conclusions}

In this paper, we presented a novel approach to service matchmaking and composition to deal with the limitations imposed by current service discovery practices. Our approach relies on representing services and user needs as graphs, which are matched using approximate graph matching based on functional graph summarization. The advocated approach expects to leverage developments in 1) functional summarization of graphs based on well understood concepts, 2) stratified graph matching based on functional summarization, 3) partial service invocation in a service description graph, and 4) service stitching, as we have outlined. The conceptual model we have introduced supports the envisioned functionalities. While the research challenges introduced are significant and ongoing, our thesis is that once these challenges are addressed, we will be in a position to discover services in an unprecedented manner, scale and at a more finer level to meet the challenges of emerging applications in large scale data integration in the age of big data.

\bibliographystyle{abbrv}
%\bibliography{bibs}

\end{document}